\documentclass[11pt]{article}
\usepackage{charter} 
\usepackage{fullpage}
\usepackage{titlesec}
\usepackage{psfrag}
\usepackage{float}
\usepackage{bbm}
\usepackage{algorithm}
\usepackage[hidelinks]{hyperref}       
\usepackage{algpseudocode}
\usepackage{url}            
\usepackage{booktabs}      
\usepackage{amsfonts}      
\usepackage{nicefrac}      
\usepackage{microtype}      
\usepackage{bm}         
\usepackage{graphicx}		
\usepackage{amsmath} 
\usepackage{cite} 
\usepackage{diagbox}
\usepackage{enumitem}
\usepackage{siunitx}
\usepackage{multirow}
\usepackage{tabularx}
\usepackage{hhline}
\usepackage{arydshln}		
\usepackage{xcolor}
\usepackage{mathtools} 
\usepackage{amsthm}			
\usepackage{amssymb}
\usepackage{setspace}
\usepackage{mathtools} 
\usepackage{amsthm}			

\newcolumntype{L}[1]{>{\raggedright\arraybackslash}p{#1}}
\newcolumntype{C}[1]{>{\centering\arraybackslash}p{#1}}
\newcolumntype{R}[1]{>{\raggedleft\arraybackslash}p{#1}}

\theoremstyle{plain} 

\def\defn{\,\coloneqq\,}

\def\sgn{{\mathsf{sgn}}}

\def\C{\mathbb{C}}
\def\R{\mathbb{R}}
\def\E{\mathbb{E}}

\def\ebm{{\bm{e}}}

\def\xbm{{\bm{x}}}
\def\zbm{{\bm{z}}}
\def\ybm{{\bm{y}}}

\def\zbm{{\bm{z}}}

\def\delbm{{\bm{\delta}}}
\def\epsilonbm{{\bm{\epsilon}}}

\def\Abm{{\bm{A}}}

\def\Fbm{{\bm{F}}}
\def\Ibm{{\bm{I}}}
\def\Mbm{{\bm{M}}}

\def\thetabm{{\bm{\theta }}}
\def\Abm{{\bm{A}}}

\def\Fbm{{\bm{F}}}
\def\Ibm{{\bm{I}}}

\def\Ncal{{\mathcal{N}}}

\def\Tsf{{\mathsf{T}}}
\def\Tsf{{\mathsf{T}}}

\usepackage[hidelinks]{hyperref}

\usepackage{upgreek}
\title{DOLPH: Diffusion Models for Phase Retrieval}

\author{
 Shirin Shoushtari$^{\dagger}$, Jiaming Liu$^{\dagger}$,~and~Ulugbek~S.~Kamilov\thanks{This material is based upon work supported by the NSF CAREER award under grant CCF-2043134.}\\
\emph{\footnotesize Computational Imaging Group (CIG), Washington University in St.\ Louis, MO, USA}\\
\emph{\footnotesize $^{\dagger}$These authors contributed equally.}
}
\date{}

\begin{document}

\maketitle
\begin{abstract}
Phase retrieval refers to the problem of recovering an image from the magnitudes of its complex-valued linear measurements. Since the problem is ill-posed, the recovery requires prior knowledge on the unknown image. We present DOLPH as a new deep model-based architecture for phase retrieval that integrates an image prior specified using a diffusion model with a nonconvex data-fidelity term for phase retrieval. Diffusion models are a recent class of deep generative models that are relatively easy to train due to their implementation as image denoisers. DOLPH reconstructs high-quality solutions by alternating data-consistency updates with the sampling step of a diffusion model. Our numerical results show the robustness of DOLPH to noise and its ability to generate several candidate solutions given a set of measurements.
\end{abstract}
%
%
\section{Introduction}
\label{sec:intro}

Phase retrieval (PR) refers to the problem of recovering phase information from noisy amplitude-only measurements. In the context of computational imaging, it is often formulated as the recovery of an unknown image $\xbm$ from the measurements
\begin{equation}
    \label{Eq:PRProblem}
    \ybm = |\Abm \xbm | + \ebm,
\end{equation}
where $\Abm \in \C^{m\times n}$ is the measurement matrix, $|\cdot|$ is the element-wise absolute value, and $\ebm$ is the noise. The popularity of PR stems from its broad applicability in microscopy~\cite{zheng2013wide}, optics~\cite{walther1963question}, astronomical imaging~\cite{fienup1987phase}, and inverse scattering~\cite{Kamilov.etal2015a}. PR is known to be challenging due to the nonlinearity of the measurements and ill-posedness of the corresponding inverse problem, necessitating algorithms that can efficiently integrate image priors. The literature on PR is vast (see the review~\cite{shechtman2015phase}) with a large number of existing methods~\cite{balan2007equi, balan2012recon, katkovnik2012phase, candes2015phase, Metzler.etal2016a, chandra2017phasepack, Metzler.etal2018}. Nonetheless, there is a strong interest in the development of PR methods that can use modern deep learning (DL) priors.

\begin{figure}[t]
\centering
\includegraphics[width=0.8\textwidth]{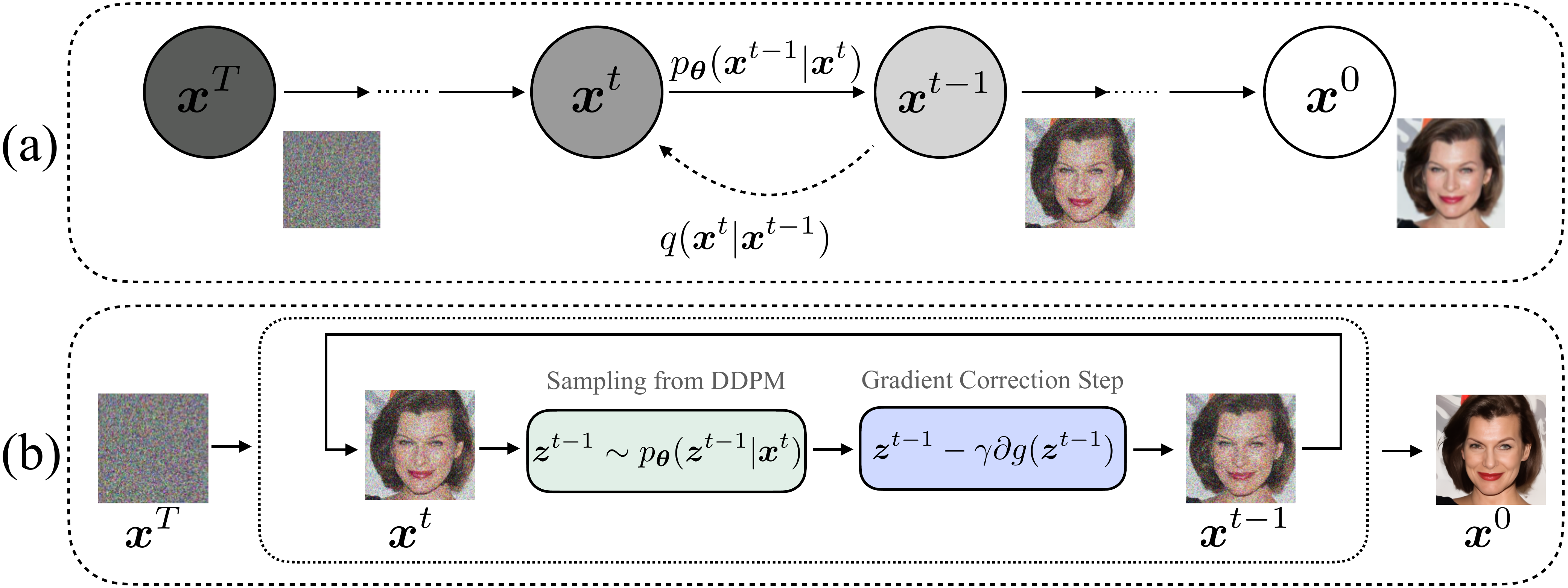}
\caption{(a) Illustration of the \emph{denoising diffusion probabilistic model (DDPM)} Markov chain that relates the distribution of high-quality images with that of the Gaussian noise. (b) DOLPH uses a pre-trained DDPM for recovering images from amplitude measurements. DOLPH can be viewed as a modified reverse diffusion process that has a gradient correction step to ensure consistency with the measured amplitudes.}
\label{Fig:schematic}
\end{figure}

\medskip\noindent
The focus of this paper is to design and validate a PR algorithm that can leverage an image prior specified by a \emph{diffusion model}. Diffusion models are a recent class of DL methods for generating high-quality images using pre-trained image denoisers~\cite{ho2020denoising, song2020improved}. The image denoiser in diffusion models can be interpreted as the gradient of the log of the image probability density function, leading to its view as a compact representation of an image distribution. This interpretation has led to the use of diffusion models as image priors in various imaging inverse problems~\cite{choi2021ilvr, song2021solving, chung2022improving, bansal2022cold}. It is also worth noting the connection between use of diffusion models in inverse problems and the plug-and-play priors (PnP) framework, which also uses image denoisers as image priors~\cite{Venkatakrishnan.etal2013, Kamilov.etal2022, kadkhodaie2021stochastic}.

\medskip\noindent
In this work, we present \emph{\textbf{d}iffusion m\textbf{o}de\textbf{l}s for \textbf{ph}ase retrieval (DOLPH)} as a new method for using diffusion models as image priors for solving PR problems of form~\eqref{Eq:PRProblem}. DOLPH combines the sampling procedure in diffusion models with the data-consistency updates ensuring that the predicted amplitudes match the measured ones in $\ybm$. Our numerical results on phase retrieval from coded diffraction patters (CDP)~\cite{Candes.etal2013} show that DOLPH can generate realistic looking images from severely noisy measurements, where the traditional approaches lead to overly smooth images.

\section{Proposed Method}
\label{sec:approach}

DOLPH builds on the sampling process of diffusion models to make it applicable to the problem in eq.~\eqref{Eq:PRProblem}. The sampling process uses a denoising convolutional neural network (CNN) pre-trained using the \emph{denoising diffusion probabilistic models (DDPM)} framework~\cite{ho2020denoising, choi2021ilvr}. DOLPH then uses the pre-trained CNN denoiser to generate samples consistent with a given set of amplitude measurements.

\subsection{Denoising diffusion probabilistic models}

\begin{algorithm}[t]
\caption{DOLPH}\label{Alg:dolph}
\begin{algorithmic}[1]
\Require $T, \ybm, \{\sigma^t\}_{t= 1}^{T}, \gamma >0$
\State $\xbm^T \sim \Ncal(0, \Ibm)$
\For{$t = T-1, \cdots, 0$}
\State $\delbm \sim \Ncal(0, \Ibm)$
\State $\zbm^{t-1} = \frac{1}{\sqrt{\alpha^t}}(\xbm^t - \frac{1 - \alpha^t}{\sqrt{1- \Bar{\alpha}^t}} \epsilon_{\thetabm}(\xbm^t, t)) + \sigma^t \delbm$

\State $\xbm^{t-1} = \zbm^{t-1} - \gamma \partial g(\zbm^{t-1})$
\EndFor
\State \textbf{return} $\xbm^0$
\end{algorithmic}
\end{algorithm}

As illustrated in Figure~\ref{Fig:schematic}(a), DDPM consists of two Markov processes: the fixed forward process and the learning-based reverse process. The forward process consists of $T \geq 1$ steps, where each step adds a Gaussian noise with a pre-designated variance, so that the at step $T$ the statistical distribution of data corresponds to the standard Gaussian distribution. Each step of the forward process can be expressed as
\begin{equation}
    \label{Eq:ForwardDiffusion}
    q(\xbm^t|\xbm^{t-1}) \defn \Ncal(\xbm^t; \sqrt{1- \beta^{t}} \xbm^{t-1}, \beta^t \Ibm),
\end{equation}
where $\Ncal(\xbm; \bm{\mu}, \bm{\Sigma})$ denotes a Gaussian probability density function with the mean vector $\bm{\mu}$ and the covariance matrix $\bm{\Sigma}$. The vectors $\xbm^1, \cdots, \xbm^T$ are the latent variables that have the same dimensions as the input image $\xbm^0$ and $\beta^t \in (0,1)$ is the variance parameter. 
By defining $\alpha^t := 1- \beta^t$ and $\Bar{\alpha}^t = \Pi_{s=1}^t \alpha^s$, the distribution of $\xbm^t$ given $\xbm^0$ can be written 
\begin{equation}
    \label{Eq:sampleNois}
    q(\xbm^t|\xbm^{0}) = \Ncal(\xbm^t; \sqrt{\Bar{\alpha}^t} \xbm_{0}, ( 1- \Bar{\alpha}^t) \Ibm).
\end{equation}
\noindent
The vector $\xbm^t$ at every step $t$ can thus be expressed as a combination of an AWGN vector $\epsilonbm \sim \Ncal (0, \Ibm)$ and $\xbm^0$~\cite{ho2020denoising}
\begin{equation}
\label{Eq:Diff2}  
\xbm^t = \sqrt{\bar{\alpha}^t}\xbm^0 + \sqrt{1-\bar{\alpha}^t} \epsilonbm.
\end{equation}

\medskip\noindent
The goal of the reverse process is to generate a clean image $\xbm^0$ given a noise vector $\xbm^T$. This is achieved by learning a CNN to reverse the Markov Chain from $\xbm^T$ to $\xbm^0$. The learning is formulated as the estimation of a parameterized Gaussian
\begin{equation}
    \label{Eq:reverse}
    p_{\thetabm}(\xbm^{t-1}|\xbm^{t}) \defn \Ncal(\xbm^{t-1}; \mu_{\thetabm}(\xbm^t, t), \sigma_{\thetabm}^2\Ibm),
\end{equation}
where $\mu_{\thetabm}(\xbm^t, t)$ and $\sigma^2_{\thetabm}$ are the learned mean vector and variance, respectively. By using the parametrization in eq.~\eqref{Eq:reverse}, the reverse update rule can be expressed as
\begin{equation}
    \label{Eq:generalsamplingrule}
    \xbm^{t-1} = \frac{1}{\sqrt{\alpha^t}}(\xbm^t - \frac{1 - \alpha^t}{\sqrt{1- \Bar{\alpha}^t}} \epsilon_{\thetabm}(\xbm^t, t)) + \sigma^t \delbm, 
\end{equation}
where $\epsilon_{\thetabm} (\xbm^t, t)$ is learned CNN that takes an AWGN corrupted input and produces the corresponding noise vector and $\delbm \sim \Ncal(0,\Ibm)$.
The CNN $\epsilon_\thetabm$ used in our implementation is learned by minimizing the following expected loss~\cite{ho2020denoising}
\begin{equation}
\label{Eq:lossDiff}
\ell (\thetabm) = \E_{\epsilonbm, \xbm, t} [\|\epsilon_\thetabm (\xbm^t, t) - \epsilonbm\|_2^2], 
\end{equation}
where $\epsilonbm \sim \Ncal (0, \Ibm)$ and $\xbm$ is the training data sample~\cite{ho2020denoising,song2020improved}. 

\begin{figure}[t]
\centering
\includegraphics[width=12cm]{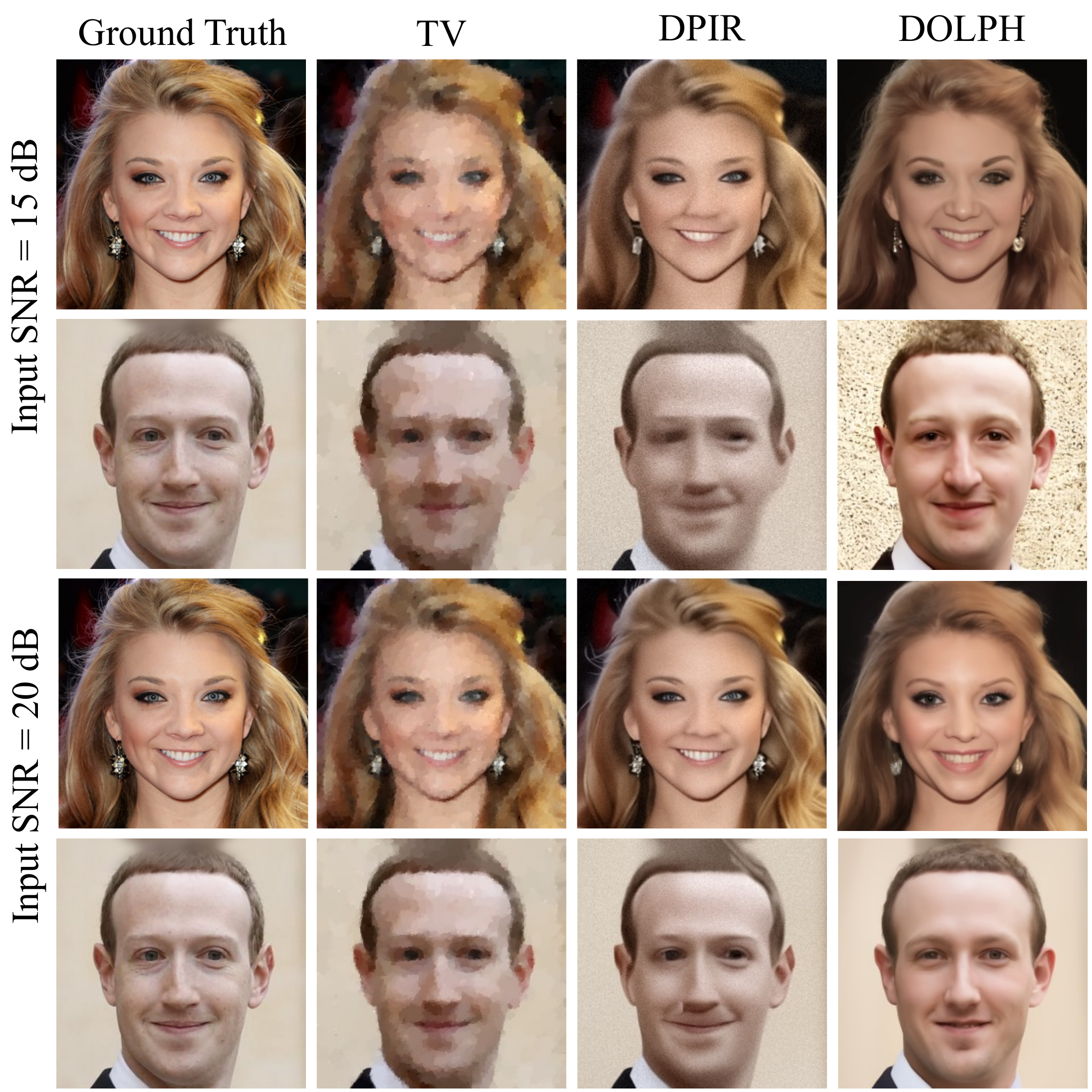}
\caption{Comparison of DOLPH against two reference methods, TV and DPIR. The first two rows correspond to the recovery from measurements contaminated with AWGN of input SNR of 15 dB, while the bottom two rows correspond to input SNR of 20 dB. Note that unlike TV and DPIR, the solutions of DOLPH are not oversmoothed.}
\label{Fig:faces}
\end{figure}

\begin{figure}[t]
\centering
\includegraphics[width=12cm]{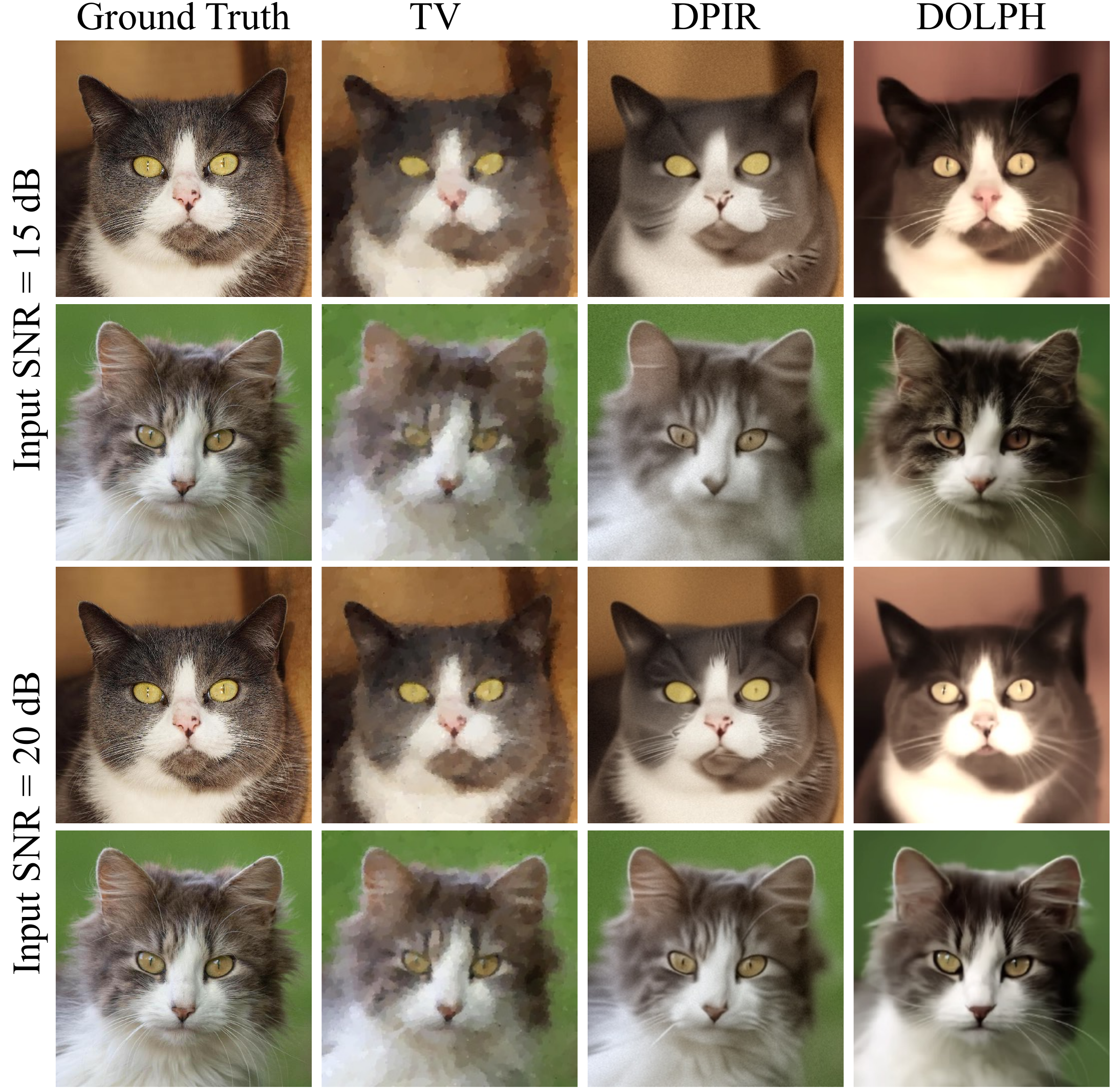}
\caption{Visual evaluation of several methods on phase retrieval from noisy intensity-only CDP measurement corresponding to 15 and 20 dB input SNR. Note that DPIR and TV fail to recover features in the image compared to DOLPH. }
\label{Fig:cats}
\end{figure}

\begin{figure}[t]
\centering
\includegraphics[width=15cm]{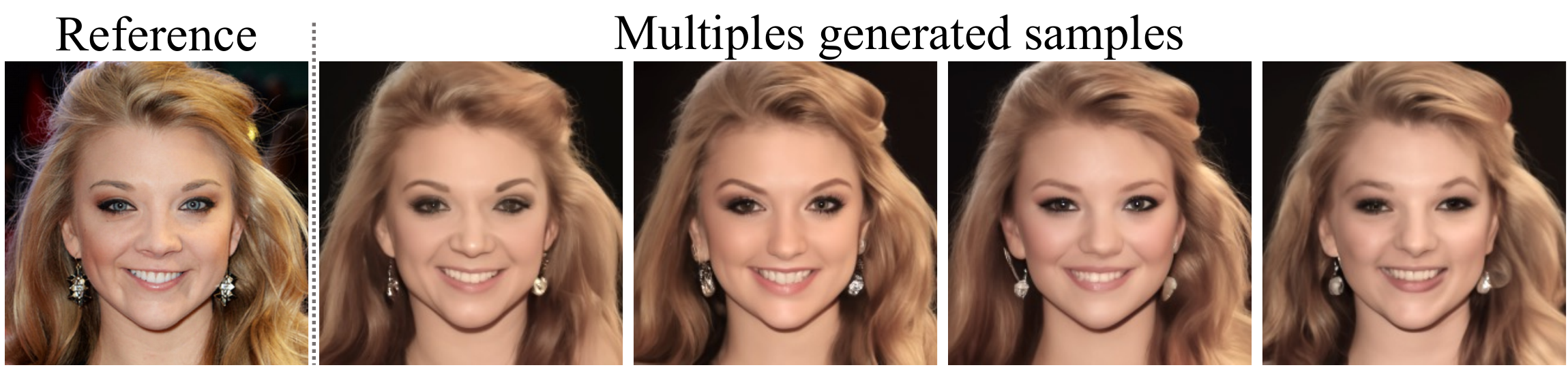}
\caption{Several generated solutions by DOLPH for the problem of phase retrieval from noisy intensity-only CDP measurements with input SNR corresponding to 15 dB.  }
\label{Fig:multisamples}
\end{figure}

\subsection{DOLPH}

The proposed DOLPH method is illustrated in Figure~\ref{Fig:schematic}(b) and summarized in Algorithm~\ref{Alg:dolph}. It consists of two key steps. The first step corresponds to the traditional DDPM sampling step
\begin{equation}
    \label{Eq:unConSampling}
    \zbm ^{t-1} \sim p_\thetabm (\zbm^{t-1}|\xbm^t)
\end{equation}
where $p_\thetabm (\xbm^{t-1}|\xbm^t)$ is defined in eq.~\eqref{Eq:reverse}, which reduces to the update rule in Line 4 of Algorithm~\ref{Alg:dolph}. The second step seeks to enforce consistency with the measurements $\ybm$ by computing 
\begin{equation}
    \label{Eq:GradLogLikelihood}
    \xbm^{t-1} = \zbm^{t-1} - \gamma \partial g(\zbm^{t-1}),
\end{equation}
where $\gamma >0 $ is the step-size parameter and $\partial g$ denotes a subgradient of the least-squares data-fidelity term
\begin{equation}
    \label{Eq:data-fidelity}
    g(\xbm) = \frac{1}{2} \|\ybm - |\Abm \xbm|\|_2^2.
\end{equation}
A subgradient of~\eqref{Eq:data-fidelity} can be expressed as
\begin{equation}
    \label{Eq:subgradient}
    \Abm^\Tsf \Abm \xbm - \Abm ^\Tsf(\ybm \cdot \sgn(\Abm\xbm)) \in \partial_\xbm \|\ybm - |\Abm\xbm|\|_2^2,
\end{equation}
where $\cdot$ denotes an element-wise product, $\sgn(\cdot)$ is the sign function, and $\partial_\xbm g$ is the subdifferential of $g$ with respect to $\xbm$.

\medskip\noindent
Following the traditional interpretation of DDPM in~\cite{ho2020denoising}, DOLPH can be related to the Langevin dynamics where $\epsilon_{\thetabm}$ is be viewed as a learned gradient of the image prior and $\partial g$ is a subgradient of the negative log likelihood. From this perspective, DOLPH is related (but distinct) from the recent work~\cite{kadkhodaie2021stochastic} that explored sampling images consistent with a learned PnP denoiser. As shown in the next section, the key benefit of using DDPM-based image denoisers as priors in DOLPH is the ability to generate high-resolution samples even in extremely noisy scenarios, where traditional PnP methods lead to overly smooth images.

\section{Numerical Results}
\label{sec:exper}

We test DOLPH on a nonconvex phase retrieval problem from \emph{coded diffraction patterns (CDP)} (similar settings were considered in~\cite{Metzler.etal2018, Wu.etal2019}). The object $\xbm \in \R^n$ is illuminated by a coherent light source. A random known phase mask modulates the light and the modulation code is denoted as $\Mbm$. In this work, each entry of $\Mbm$ is drawn uniformly from the unit circle in the complex plane. The light goes through the far-field Fraunhofer diffraction and a camera measures its intensity $\ybm \in \R_+^m$. Since the Fraunhofer diffraction can be modeled by a Fourier Transform, the data-fidelity term of this phase reconstruction problem can be formulated as
\begin{equation}
    \label{Eq:phaseRecon}
    g(\xbm) = \frac{1}{2} \|\ybm - |\Fbm \Mbm \xbm|\|_2^2,
\end{equation}
where $\Fbm$ is 2D discrete Fast Fourier Transform (FFT). 

\begin{figure*}[t]
\centering
\includegraphics[width=16.5cm]{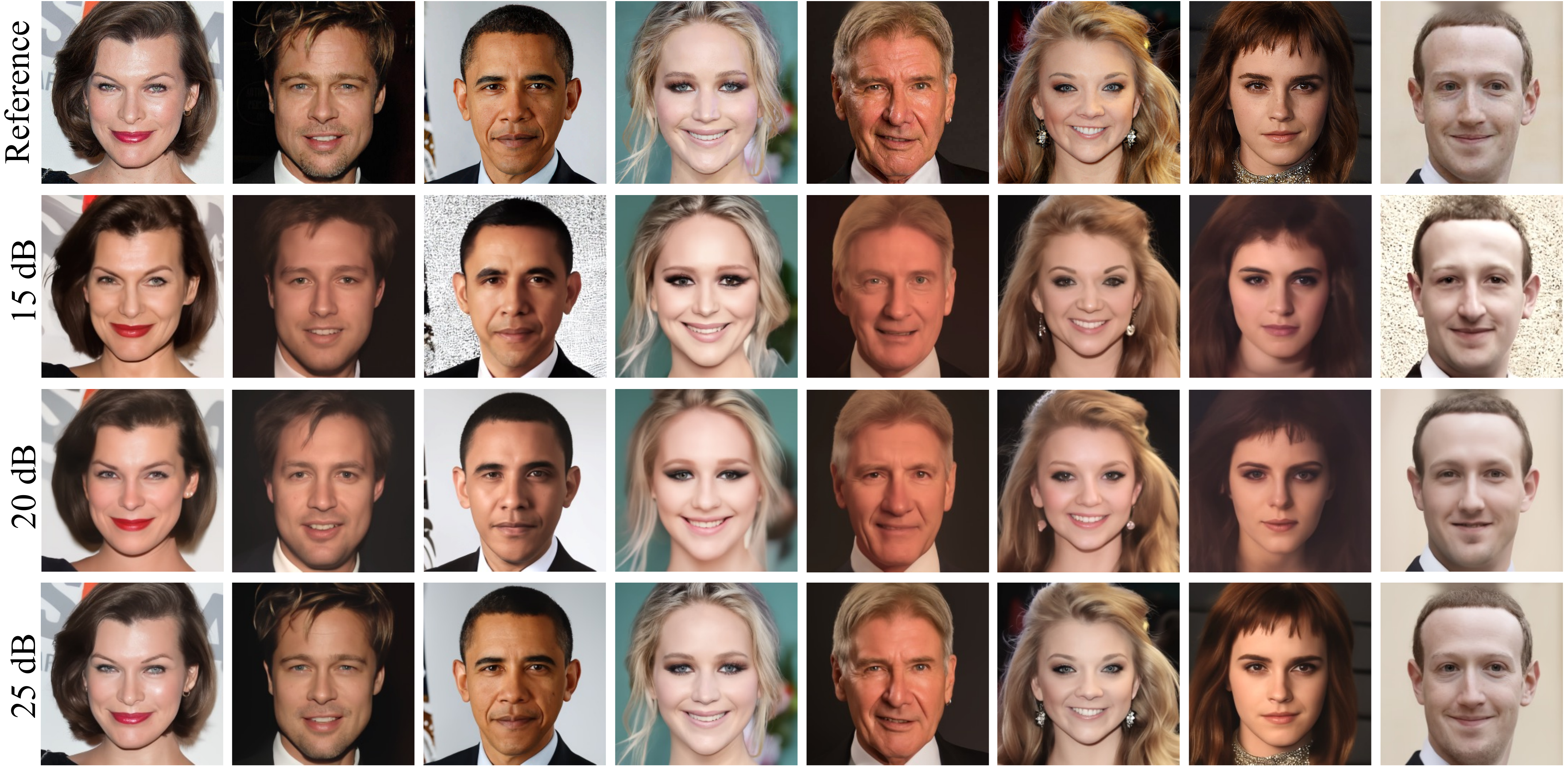}
\caption{Several solutions by generated by DOLPH for the problem of phase retrieval from noisy intensity-only CDP measurements with input SNR corresponding to 15, 20 and 25 dB. Note the improvement in the quality of samples as the amount of noise decreases. At 25 dB input SNR, the solutions of DOLPH nearly match the groundtruth images in the first row.}
\label{Fig:multisamp}
\end{figure*}

\medskip\noindent
We use eight face images collected from the web (see Figure~\ref{Fig:multisamp}) and four cat images from the dataset~\cite{choi2020stargan} as test data. We recover 256 $\times$ 256 color images, with simulated CDP measurements corrupted by AWGN corresponding to 15, 20, and 25 dB input signal-to-noise ratio (SNR). For reference, we include two well-known baseline methods for PR problem, including TV~\cite{Rudin.etal1992} and DPIR~\cite{zhang2021plug}. TV is an iterative method that does not require any training, while DPIR is a variant of plug-and-play priors (PnP) that uses a residual U-Net architecture from~\cite{zhang2021plug} as the image denoiser. The regularization parameters for both TV and DPIR were optimized for every individual image for the best imaging quality. We used an unconditional DDPM pre-trained on FFHQ~\cite{karras2019style}, MetFaces~\cite{karras2020training}, and AFHQ~\cite{choi2020stargan} as an image prior for DOLPH\footnote{\url{https://github.com/jychoi118/ilvr_adm}}.

\medskip\noindent
Figures~\ref{Fig:faces} and~\ref{Fig:cats} present a visual comparison of images obtained using DOLPH, TV, and DPIR at input SNR values corresponding to 15 and 20 dB. The visual comparison suggests that TV and DPIR generate overly smooth images at high noise levels, while DOLPH solutions always preserve high-resolution features. As can be seen on data corrupted with noise corresponding to 15 dB input SNR, DOLPH images can look realistically sharp even when they return a face different from groundtruth. While this results in lower reconstruction SNR for DOLPH compared to both TV and DPIR, DOLPH can be used to generate multiple solutions consistent with a given measurement vector. Figure~\ref{Fig:multisamples} illustrates several candidates solution for phase retrieval problem from noisy intensity-only CDP measurements corresponding to 15 dB input SNR. Note the quality of each generated solutions by DOLPH in comparison with DPIR solution. DOLPH manages to recover sharp, high-resolution images despite the severity of the noise corruption.

Figures~\ref{Fig:multisamp} presents examples for all the images in our training set for each noise level. Note how at 25 dB input SNR, the solutions of DOLPH (shown in the fourth row of the figure) nearly match the groundtruth images (shown in the first row of the figure).

\section{Conclusion}
This work presented DOLPH as a new deep model-based architecture for phase retrieval. DOLPH leverage a diffusion model as an image prior with a nonconvex data-fidelity term. We compared the performance of DOLPH with two existing methods, TV and DPIR. Our results show that under high amount of noise, TV and DPIR result in overly smooth images, while DOLPH can still generate more realistic images. Additionally, DOLPH can be used to generate multiple images with a given measurement vector. Our results suggest to the potential of diffusion models to act as image priors in nonconvex image reconstruction problems. 
\label{con}


\end{document}